\documentclass{appolb}
\usepackage{epsfig}
\usepackage{amsmath, amsthm}
\usepackage{amsfonts}

\begin{document}
\eqsec  
\title{CAN CAUSAL DYNAMICAL TRIANGULATIONS PROBE FACTOR-ORDERING ISSUES?
\thanks{Presented at the XLIX Cracow School of Theoretical Physics, ``Non-perturbative Gravity and Quantum Chromodynamics,''
Zakopane, Poland, May 31$-$June 10, 2009.}%
}
\author{R L Maitra
\address{Institute for Theoretical Physics, Utrecht University, Leuvenlaan 4, 3584 CE Utrecht, The Netherlands}
}
\maketitle
\begin{abstract}
The causal dynamical triangulations (CDT) program has for the first time allowed for path-integral computation of correlation
functions in full general relativity without symmetry reductions and taking into account Lorentzian signature.  One of the most exciting recent results in CDT is the strong agreement of these computations with (minisuperspace) path integral calculations in quantum cosmology.  Herein I will describe my current project to compute minisuperspace (Friedman-Robertson-Walker) path integrals with a range of different measures corresponding to various factor orderings of the Friedman-Robertson-Walker Hamiltonian.  The aim is to compare with CDT results and ask whether CDT can shed light on factor-ordering
ambiguities in quantum cosmology models.
\end{abstract}
  
\section{Introduction}
In every approach to quantum gravity, persistent challenges from diffeomorphism invariance rear their heads, with forms varying according to method of quantization.  One particularly unregenerate challenge comes from the fact that the kinetic term in the Hamiltonian of a diffeomorphism-invariant theory contains products of position and momentum variables.  In a canonical approach, this leads to factor ordering ambiguities because the kinetic term in the quantum Hamiltonian will contain products of noncommuting operators, while in a path integral setting it results in a corresponding indeterminacy in the definition of the functional integral measure.

Before even tackling the definition of a path integral measure for general relativity, however, we encounter a more immediate difficulty engendered by the sheer enormity of the set of paths to be integrated over.  The space of histories is through superspace, the set of all possible spatial geometries (3-metrics modulo diffeomorphisms).  In order to make sense of the integration, let alone perform practical computations, one is forced to restrict the class of geometries considered by imposing one or another simplifying assumption.  Most common are two fundamentally different approaches to reducing the space of allowed geometries.  

Chronologically first came the method of reducing superspace to a ``minisuperspace'' by restricting to geometries so symmetric that only finitely many degrees of freedom remain \cite{DeWittMini, MisnerMini}.  With the obvious convenience of reducing from a quantum field theory to an effectively quantum mechanical system, one can consider propagators and correlators for quantities of cosmological interest such as the scale factor.  These advantages are alloyed by the danger of constructing a quantum theory in which artifacts of the symmetry reduction overshadow genuine quantum gravitational physics; however, many issues to be faced in the quantization of full general relativity are preserved in these quantum cosmological models.  As a mixed blessing, the twin problem of factor ordering and path integral measure persists.  Because of quantum cosmology's great potential power and utility, one would dearly like to know how far to trust it as a model problem for full quantum gravity.

Another, newer approach to the reduction of superspace, the method of causal dynamical triangulations (CDT), imposes no symmetries but instead constructs a regularized path integral by restricting to piecewise flat geometries \cite{ALFirst, AJLFirst, AJL}.  Spatial slices are constructed of 3-simplices, and joined to adjacent spatial slices by edges which complete the whole into a 4-d simplicial manifold.  These joining timelike edges are required to connect in a fashion which preserves the causal structure of the slicing (see e.g. \cite{AJL}), so that although the path integral performed is Euclidean, only causally well-behaved histories are summed over.  

The Euclidean Einstein-Hilbert action is replaced in CDT by the Euclidean Regge action; the path integral by a sum over triangulations: 
\begin{align*}
\int {\cal D} \left[ g \right] e^{ - S_E^{EH} \left[ g \right]} \quad \to \quad\sum_{\cal T}  \frac{1}{C_{\cal T}} e^{ - S_E^{{\rm Regge}} \left( {\cal T} \right)} 
\end{align*}
where ${\cal T}$ is a causally well-behaved triangulation of spacetime (taken to be topologically $I \times S^3$).  The quantity $C_{\cal T}$ is equal to the order of automorphism group of the triangulation ${\cal T}$, and the $1/C_{\cal T}$ factor in the summation is known as the discrete measure because it weights highly symmetric geometries with a lower probability.  In practice, the sum over triangulations is approximated using a Monte Carlo simulation which generates a set of independent histories.

Because the CDT simulation deals directly with (triangulated) geometries and computes discrete averages rather than continuum integrals, this choice of discrete measure to employ in the summation has a far more transparent meaning than is possible in quantum cosmological models.  

Recently, computations have become possible which allow comparison of results from CDT with those from the semiclassical or quadratic approximations in the quantized Friedman-Robertson-Walker (FRW) cosmological model \cite{AJLSemiFirst, AGJLPlanckBirth, AGJL}.  The semiclassical state in FRW can be recovered from CDT simulations, and by computing the covariance matrix of the (discrete) spatial 3-volume, one can even observe CDT realizing quantum fluctuations in 3-volume.  To an impressive degree, these fluctuations match those given by the quadratic approximation of quantized FRW.  

It is in this second comparison, of the CDT covariance matrix with the FRW quadratic approximation for volume fluctuations, that I show how the signature of factor ordering can be detected.  By performing an FRW quadratic approximation analogous to that in \cite{AGJLPlanckBirth} and \cite{AGJL} but with varying choice of factor ordering and therefore varying path integral measure, I set up a range of possible path integral expressions with the goal of performing each and comparing to data from CDT to determine which quantum cosmological path integral measure most closely approximates path integrals performed in a non-symmetry-reduced manner.  In this way, CDT may be able to help us decide which path integral measure for FRW quantum cosmology most faithfully represents the full quantum gravitational path integral with all modes except the global scale factor integrated out.

The plan of the paper is as follows.  Section \ref{CDT-FRW} summarizes the calculation in \cite{AGJLPlanckBirth} and \cite{AGJL} comparing CDT's 3-volume covariance matrix with the quadratic approximation for quantum fluctuations in the FRW model, and points out the stage at which path-integral measure dependence enters in.  Section \ref{WKB} adopts a 2-parameter family of factor orderings for FRW quantum cosmology first studied by Steigl and Hinterleitner in \cite{SteiglHinterleitner}, and uses a WKB-style approximation for the solution to the FRW Wheeler-DeWitt equation to estimate the computational effects at quadratic level of using different factor orderings.  In Section \ref{pathint}, I demonstrate that path integral expressions corresponding to all the Steigl-Hinterleitner factor orderings can be obtained in relation to the known path integral expression (due to DeWitt \cite{DeWitt} and Parker \cite{Parker}) for a particle in a potential on a curved background.  Finally, Section \ref{discussion} discusses the results and outlines directions of future research.  

\section{CDT and the FRW model}
\label{CDT-FRW}

In the Monte Carlo simulations which provide the computational backbone of CDT, measuring the covariance matrix for spatial volume is a conceptually straightforward affair.  Using the $K$ independent histories generated by the simulation, one keeps track of the spatial volume $N_3^{\left( k \right)} \left( i \right)$ (the number of 3-simplices) for each history $k$, at each timestep $i$.  Following the notation in \cite{AGJL}, the expectation value for spatial volume is approximated by a simple average over the histories at hand:
\begin{eqnarray*}
\overline N _3 \left( i \right) \equiv \langle N_3 \left( i \right) \rangle \cong \frac{1}{K}\sum\limits_k N_3^{\left( k \right)} \left( i \right). 
\end{eqnarray*}
The covariance matrix for spatial volume is now given by
\begin{eqnarray}
\label{covmat}
C\left( i,j \right) \cong \frac{1}{K} \sum_k \left( N_3^{\left( k \right)} \left( i \right) - \overline N _3 \left( i \right) \right) \left( N_3^{\left( k \right)} \left( j \right) - \overline N _3 \left( j \right) \right).
\end{eqnarray}

To compare this covariance matrix with results from FRW quantum cosmology, one must first write the Euclidean signature (Wick- and conformal-rotated) FRW Lagrangian in terms of the spatial volume rather than the scale factor:
\begin{eqnarray}
\label{vollag}
L\left[ {V_3 \left( s \right)} \right] = \frac{c_1 }{V_3 \left( s \right)}\left( \frac{dV_3 \left( s \right)}{ds} \right)^2  + c_2 V_3 ^{1/3} \left( s \right) - \lambda V_3 \left( s \right).
\end{eqnarray}
Denoting the quantum fluctuations in spatial volume by $x(t) \equiv V_3(t) - V_3^{cl}(t),$ where $V_3^{cl}(t)$ is the classical solution, the correlator for the volume fluctuations can be approximated by the usual WKB expansion, to quadratic order:
\begin{eqnarray}
\label{quadcorr}
\langle x \left( t \right) x\left( t' \right) \rangle & \approx & e^{ - \frac{S \left[ V_3^{cl} \right]}{\hbar }} \cdot \int {\cal D} x \left( s \right) x \left( t \right) x \left( t' \right) e^{ - \frac{1}{2 \hbar } \iint ds \ ds' \ x \left( s \right) M \left( s,s' \right) x \left( s' \right)} \nonumber \\ 
\ &=& e^{ - \frac{S \left[ V_3^{cl} \right]}{\hbar }} \cdot M^{ - 1} \left( t,t' \right),
\end{eqnarray}
according to standard path integral computations.

However the final equality in (\ref{quadcorr}) depends upon the assumption that the path integral is Gaussian; i.e. that taken together, the expression ${\cal D} x \left( s \right) e^{ - \frac{1}{2 \hbar } \iint ds \ ds' \ x \left( s \right) M \left( s,s' \right) x \left( s' \right)}$ is a functional Gaussian measure.  Because the kinetic term in the Lagrangian (\ref{vollag}) involves both $V_3$ and its time derivative, the situation is in fact complicated by ambiguity in factor ordering and hence choice of path integral measure.  Here we receive our first hint that perhaps even this simple quadratic approximation computation may be affected by the factor ordering problem.  Far from being a misfortune, this added complexity in the quadratic approximation is our opportunity to probe the factor-ordering problem in quantum cosmology with currently accessible computational tools.

The goal of this paper is to revisit the computation (\ref{quadcorr}), taking into account variation in path integral measure resulting from varying choice of factor ordering in the quantum FRW Hamiltonian.  For ease of contact with other approaches to the factor ordering problem, the remainder of the paper considers the FRW Hamiltonian in terms of the scale factor rather than spatial volume, to which it is related simply by $V_3 \cong a^3$.

I show that by ranging over the two-parameter family of factor orderings for the quantum FRW Hamiltonian introduced in \cite{SteiglHinterleitner}, the correlator computed in (\ref{quadcorr}) will vary in a well-defined and quantifiable fashion.  Ultimately the aim is to compare the computed correlators for each factor ordering with CDT data and determine a best fit.

Before proceeding, a simple WKB approximation to the solution of the Wheeler-DeWitt equation for varying factor ordering gives us an assessment of the extent to which we may expect results at the quadratic level to depend on factor ordering.

\section{How much difference does choice of measure make?}
\label{WKB}

Classically, the Hamiltonian for the FRW universe is given by 
\begin{eqnarray}
\label{ham}
H = \frac{1}{2}\left[-\frac{p_{a}^{2}}{a} - a + \frac{\Lambda}{3}a^{3} \right],
\end{eqnarray}
where as usual $\Lambda$ is the cosmological constant.  In \cite{SteiglHinterleitner}, the form of the quantized Hamiltonian is allowed to range over a 2-parameter family of factor orderings, so that
\begin{eqnarray}
\label{qham}
\hat H &=& \frac{1}{2}\left[\hbar^{2} a^{-i} \partial_{a} a^{-j} \partial_{a} a^{-k} - a + \frac{\Lambda}{3}a^{3} \right] \nonumber \\
\ &=& \frac{\hbar^{2}}{2} a^{-i} \partial_{a} a^{-j} \partial_{a} a^{-k} + V(a),
\end{eqnarray}
where $i+j+k=1$ and $V(a) = \frac{1}{2}\left(-a + \frac{\Lambda}{3}a^{3} \right)$.  This Hamiltonian is self-adjoint on the Hilbert space of states $L^{2}\left( \mathbb{R}^{+}, a^{i-k}da \right)$.

Two special cases in the family of Steigl-Hinterleitner factor orderings are worthy of note:  first, by taking $i=k$, we obtain a 1-parameter family of symmetric factor orderings for which the Hamiltonian (\ref{qham}) is self-adjoint with respect to the ordinary Lebesgue measure $da$.  Alternatively, by considering $i=j=1/2$, $k=0$, we can make the kinetic term of (\ref{qham}) into a covariant Laplace-Beltrami derivative operator
\begin{align}
\Delta_{LB} \equiv - \left( dd^{*} + d^{*}d \right) = \frac{1}{\sqrt{ \left| g \right| }}\partial_\alpha  \sqrt {\left| g \right|} g^{\alpha \beta } \partial _\beta,
\end{align}
which for a 1-dimensional metric $g(a) = (a)$ is given by
\begin{align}
a^{ - 1/2} \partial_a a^{ - 1/2} \partial _a.
\end{align}
This choice $i=j=1/2$, which we can call the ``Laplace-Beltrami" factor ordering, is the one we would choose in order to regard the Hamiltonian (\ref{ham}) as effectively that of a particle in a potential $V(a)$ on a curved background $g(a) = (a)$.

The quantized Hamiltonian (\ref{qham}) yields the Wheeler-DeWitt equation
\begin{align}
\label{WDW}
\hbar ^2 a^{ - 1} \partial _a^2 \psi \left( a \right) & + \hbar ^2 \left( { - k + i - 1} \right)a^{ - 2} \partial _a \psi \left( a \right) \nonumber \\ 
& + \left[ {\hbar ^2 k\left( { - i + 2} \right)a^{ - 3}  - a + \frac{\Lambda }{3}a^3 } \right]\psi \left( a \right) = 0,
\end{align}
for which we can construct a WKB-style series solution of the form
\begin{eqnarray}
\label{WKBsol}
\psi \left( a \right) = \exp \left\{ - \frac{S\left( a \right)}{\hbar } \right\}\sum_{n = 0}^\infty {\hbar^n } \varphi_n \left( a \right).
\end{eqnarray}
In standard path integral treatments, one uses arguments from the method of steepest descent to assert that in the expansion of a path integral about its classical solution
\begin{align}
\int {\cal D} x \exp \left\{ - \frac{1}{\hbar} \left( S_{cl} + \frac{1}{2!} \iint \frac{\delta^{2}S}{\delta x_{cl}^{2}} \left( x - x_{cl} \right)^{2} + \frac{1}{3!} \iiint \frac{\delta^{3} S}{\delta x_{cl}^{3}} \left(x - x_{cl} \right)^{3} + \dots \right) \right\},
\end{align}
terms of order $\left( x - x_{cl} \right)^{3}$ and higher in the exponent yield terms in a summation expansion of order $\hbar$ and higher.  Using this reasoning, the quadratic approximation for the path integral is equivalent to the limit $\hbar \to 0$ and thus to the truncation of the WKB series solution of the Wheeler-DeWitt equation from the form (\ref{WKBsol}) to
\begin{align}
\label{WKB1}
\varphi_0 \left( x \right)\exp \left\{ - \frac{S\left( x \right)}{\hbar } \right\}.
\end{align}

The arguments necessary to make such an equivalence of approximations precise would depend on the definition of the path integral measure.  In our case, this is precisely what is in question, so truncating the WKB series solution from (\ref{WKBsol}) to (\ref{WKB1}) must not be taken as anything more than a very rough heuristic guide to what we may find upon performing a quadratic approximation of the path integral.  Still, we shall find that solving for (\ref{WKBsol}) up to $\varphi_{0}$ proves to be an illuminating exercise.

Inserting (\ref{WKBsol}) into the Wheeler-DeWitt equation (\ref{WDW}) yields, to second order,
\begin{subequations}
\begin{eqnarray}
S\left( a \right) = \frac{1}{\Lambda }\left( {1 - \frac{\Lambda }{3}a^2 } \right)^{\frac{3}{2}}  - \frac{1}{\Lambda } \label{hamprin} \\ \varphi _0 \left( a \right) = K\left| {1 - \frac{\Lambda }{3}a^2 } \right|^{{\rm{ }} - {\rm{ }}\frac{1}{4}}  \cdot a^{\frac{k - i}{2}} \label{h0},
\end{eqnarray}
\end{subequations}
where coefficient $K$ in (\ref{h0}) is an arbitrary constant.  As expected, (\ref{hamprin}) is the solution to the imaginary-time Hamilton-Jacobi equation.  Hence $\exp \left[ -S(a) \right]$ is the semiclassical state and is independent of factor ordering since (\ref{hamprin}) does not involve $i, j,$ or $k$.

As a clue to whether the quadratic approximation may carry the signature of factor ordering, the important feature of (\ref{h0}) is the factor $a^\frac{k-i}{2}$, which indicates that at quadratic level we may hope to measure the deviation in factor ordering from the symmetric case $i=k$.  Emboldened by this result on the canonical side, we turn to path integral computations.

\section{Path integrals for Steigl-Hinterleitner orderings}
\label{pathint}
To compare our results with those from CDT simulations, we must consider correlators in the path integral formalism, in particular correlators for functions of the FRW scale factor.  As in CDT, proper time will be used throughout for the FRW model.  However to define path integrals with various measures in terms of time-slicing, it is more natural to deal with propagators than correlators.  This mismatch is circumvented by the simple observation that correlators can always be broken down into expressions involving propagators:
\begin{flalign}
\label{corr-prop}
\langle a'', t'' | f \left( \hat a \left( t_2 \right) \right) & f \left( \hat a \left( t_1  \right) \right)  | a',t' \rangle_{ijk} = \nonumber \\ 
& \int_{0}^{\infty}  a_2^{i - k} da_2 \ f \left( a_2 \right) \langle a'',t'' | a_2 ,t_2 \rangle_{ijk} \ \times \nonumber \\ 
& \ \ \left[ \int_{0}^{\infty} a_1^{i - k} da_1 f \left( a_1 \right) \langle a_2 ,t_2 | a_1, t_1 \rangle_{ijk} \langle a_1, t_1 | a', t' \rangle_{ijk} \right],
\end{flalign}
by using the appropriate resolution of identity.  The subscript $ijk$ labels the correlators and propagators as those corresponding to a given choice of Steigl-Hinterleitner factor ordering of the FRW Hamiltonian.  

Since correlators can be expressed in terms of propagators, the focus can shift to finding a path integral expression for the propagator $K_{ijk} \left( a,t;a',t' \right)$ satisfying
\begin{align}
& \left( \frac{\partial } {\partial t} - \frac{\hbar ^2 }{2} a^{-i} \partial_a a^{-j} \partial_a a^{-k} - V(a) \right) K_{ijk} \left( a,t;a',t' \right) = 0 , \ \ a \ne a' \nonumber \\ 
& \lim_{t \searrow 0} K_{ijk} \left( a,t;a',t' \right) = \left( a' \right)^{k-i} \delta \left( a - a' \right).
\end{align}

For the case of the Laplace-Beltrami factor ordering, the path integral expression for this propagator is known \cite{DeWitt, Parker}.  It is the propagator for a particle moving in a potential $V(a)$ on a curved background $g(a)=(a)$:
\begin{align}
\label{DeWitt}
 K_{LB} \left( a'',t'';a',t' \right) &= \int_{ {\cal C} \left\{ a'', t'' | a', t' \right\}} {\cal D} \left[ a^{\frac{1}{2}} \left( t \right) a \left( t \right) \right] \exp \left\{  - \frac{1}{\hbar} \int_{t'}^{t''} dt \left[ \frac{1}{2} a \dot a^2 - V(a) \right]  \right\} \nonumber \\ 
& \equiv \lim_{N \to \infty } \frac{1}{\left( 2\pi \varepsilon \hbar \right)^{N/2} } \prod_{n = 1}^{N - 1} \int a_n^{\frac{1}{2}} da_n \nonumber \\ 
& \ \ \ \ \ \times \exp \left\{ - \frac{1}{\hbar } \sum_{n = 1}^N \left[ \frac{1}{2\varepsilon } a_{n-1} \left( a_n - a_{n-1} \right)^2 - \varepsilon V\left( {a_{n - 1} } \right) \right] \right\}.
\end{align}
In the remainder of this section, I show that path integral expressions for all other Steigl-Hinterleitner factor-ordered FRW Hamiltonians can be obtained in relation to (\ref{DeWitt}).  

For any ordering $(i,k)$ in the 2-parameter family of Steigl-Hinterleitner orderings, a two-step process relates the $(i,k)$ propagator to (\ref{DeWitt}).  First, the propagator for any ordering belonging to the line $i + k = 1/2$ can be related to (\ref{DeWitt}) by convolution with appropriately chosen Green's functions.  Next, the propagator for an ordering belonging to any perpendicular line $i - k = c$ ($c = \rm{constant}$) can be related to a propagator whose kinetic term is ordered according to $i - k = c, \ i + k = 1/2$, and whose potential term differs from $V(a)$ by the addition of a quantum potential proportional to $\hbar^{2}$.  Using the convolution result described above for orderings on the line $i + k = 1/2$, this new propagator with added quantum potential can then be related to the propagator (\ref{DeWitt}) with a corresponding added quantum potential.  These results are described in detail below.

\subsection{Orderings with $i + k = 1/2$}
\label{conv}
For orderings lying along the line $i + k = 1/2$, the quantum FRW Hamiltonian is
\begin{align}
\hat H = \frac{\hbar ^2 }{2}\hat a^{k - 1/2} \partial _a \hat a^{ - 1/2} \partial _a \hat a^{ - k}  + V(\hat a),
\end{align}
so we need a propagator satisfying
\begin{subequations}
\begin{eqnarray}
\left( \frac{\partial }{\partial t} - \hat H \right)K_{ijk} \left( a'',t'';a',t' \right) = 0, \quad a'' \ne a' \label{Schro} \\
\lim_{t \searrow 0} K_{ijk} \left( a'',t'';a',t' \right) = \left( a' \right)^{2k - 1/2} \delta \left( a'' - a' \right).
\end{eqnarray}
\end{subequations}

Note that we can write the Schr{\"o}dinger operator in (\ref{Schro}) as
\begin{align}
\left( \frac{\partial }{\partial t} - \hat H \right) = \hat a^k \left( \frac{\partial }{\partial t} - \hat H_{LB} \right) \hat a^{ - k},
\end{align}
where $\hat H_{LB}$ is the quantum FRW Hamiltonian with Laplace-Beltrami factor ordering.  Hence by convolving the propagator $K_{LB}$ for the Laplace-Beltrami Schr{\"o}dinger operator with the Green's functions $a^{k} \delta (a-a')$ and $a^{-k} \delta (a-a')$ for the multiplication operators $\hat a^{-k}$ and $\hat a^{k}$, and converting to the correct measure $a^{1/2 - 2k} da$, we obtain the propagator $K_{ijk}$:
\begin{align}
 K_{ijk} \left( a'',t'';a',t' \right) &= \left( a' \right)^{2k - 1/2} \left( a' \right)^{1/2} \left[ \left( a'' \right)^k K_{LB} \left( a'',t'';a',t' \right)\left( a' \right)^{ - k}  \right] \nonumber \\ 
&= \left( a'a'' \right)^k K_{LB} \left( a'',t'';a',t' \right).
\end{align}

\subsection{Orderings with $i - k = c$}
All orderings lying outside the line $i + k = 1/2$ are dealt with according to which perpendicular line $i - k = c$ they occupy.  First, rewrite the quantum Hamiltonian in terms of a new effective kinetic term and a quantum potential added to $V(a)$:
\begin{align}
\label{QP}
\hat H &= \frac{\hbar ^2 }{2} a^{ - k - c} \partial _a a^{2k + c - 1} \partial _a a^{ - k}  + V(a) \nonumber \\ 
&= \frac{{\hbar ^2 }}{2}\left[ {a^{ - 1} \partial _a^2  - \left( {1 - c} \right)a^{ - 2} \partial _a  + k\left( {2 - c - k} \right)a^{ - 3} } \right] + V(a) \nonumber \\ 
&= \frac{\hbar ^2 }{2} \left[ a^{ - 1/2 \left( 1/2 + c \right) } \partial _a a^{ - 1/2} \partial _a a^{ - 1/2 \left( 1/2 - c \right) }  \right] \nonumber \\ 
& \quad + \frac{\hbar ^2 }{2}\left[ k \left( 2 - c - k \right) - \frac{1}{4} \left( \frac{1}{2} - c \right) \left( \frac{7}{2} - c \right) \right] a^{-3} + V(a)
\end{align}
Notice that the effective kinetic term in the second to last line is ordered according to a different Steigl-Hinterleitner ordering having $i^{'} = (1/2 + c)/2, \ j^{'} = 1/2, k^{'} = (1/2 - c)/2$.  This ordering satisfies $i^{'} + k^{'} = 1/2$, so now the convolution procedure of Section \ref{conv} applies, with the new effective potential on the last line of (\ref{QP}) instead of simply $V(a)$.  Thus we obtain
\begin{align}
\label{coup}
K_{ijk} & \left( a'', t'' ; a' , t' \right) = \left( a'a'' \right)^{1/4 + \frac{k - i}{2}} \int_{{\cal C}\left\{ a'',t'' | a',t' \right\}} {\cal D} \left[ a^{\frac{1}{2}} \left( t \right) a \left( t \right) \right] \times \nonumber \\
& \exp \left\{ - \frac{1}{\hbar } \int_{t'}^{t''} dt \left[ \frac{1}{2} a \dot a^2  - \left( V(a) + \frac{\hbar ^2 }{2} p \left( k,c \right) a^{ - 3} \right) \right] \right\}, \\
& \quad p \left( k,c \right) = k \left( 2 - c - k \right) - \frac{1}{4} \left( \frac{1}{2} - c \right) \left( \frac{7}{2} - c \right). \nonumber
\end{align}

The expression (\ref{coup}) gives the path integral expression for the propagator corresponding to any Steigl-Hinterleiter ordering, as a well-defined deviation from the Laplace-Beltrami case.  The deviation from the Laplace-Beltrami propagator enters in two ways:  through the quantum potential measuring the ordering's distance from the line $i + k = 1/2$, and through the prefactor measuring the ordering's degree of asymmetry.

Reversing the breakdown of correlators into propagators given by (\ref{corr-prop}), we can use (\ref{coup}) to express any Steigl-Hinterleitner correlator for a function $f(a)$ of the scale factor in terms of the Laplace-Beltrami correlator for the same function:
\begin{align}
\label{prop}
 \langle a'',t'' | &f \left( \hat a \left( t_2 \right) \right) f \left( \hat a \left( t_1 \right) \right) | a',t' \rangle _{ijk} \nonumber \\
 & = \left( a''a' \right)^{\frac{1}{4} + \frac{k - i}{2}} \langle a'',t'' | f \left( \hat a \left( t_2 \right) \right) f \left( \hat a \left( t_1 \right) \right) | a',t' \rangle _{LB,QP}, 
 \end{align}
where the subscript $LB,QP$ denotes the fact that the correlator on the right-hand side is taken with the Laplace-Beltrami factor ordering, and with the added quantum potential.

\section{Discussion}
\label{discussion}
As a confirmation of the WKB prediction in Section \ref{WKB}, notice that the prefactor
\begin{align*}
\left( a''a' \right)^{\frac{1}{4} + \frac{k - i}{2}}
\end{align*}
in (\ref{coup}) and (\ref{prop}) measures the effect on propagators and correlators of the factor ordering's degree of asymmetry, in a manner closely resembling that of the ordering-dependence in the first term of the WKB fluctuation factor (\ref{h0}):
\begin{align}
\varphi _0 \left( a \right) = K\left| {1 - \frac{\Lambda }{3}a^2 } \right|^{{\rm{ }} - {\rm{ }}\frac{1}{4}}  \cdot a^{\frac{k - i}{2}}
\end{align}
The obvious next step is to compute the path integral in (\ref{coup}) to quadratic order, expecting based on WKB analysis that ordering dependence will persist.  To compute the quadratic approximation to the path integral, techniques such as those applied in \cite{BekPark} and references therein may prove useful.

Having computed the path integrals in (\ref{coup}) and (\ref{prop}) to quadratic level, the goal is of course to compare with the CDT data for volume fluctuations described in Section \ref{CDT-FRW}.  This will necessitate the conversion from results in terms of the scale factor to results in terms of spatial volume.  However, this is a straightforward affair which could be approached either by computing correlators in terms of $a^{3}$ or by recreating the above derivations for the Lagrangian (\ref{vollag}).

Once we have computed the FRW volume fluctuation correlator for each path integral measure $\left[ \mathcal{D}a \right]_{(i,k)}$ in the 2-parameter family, we can compare against the CDT covariance matrix (\ref{covmat}) and determine the $(i,k)$ which yields a best fit.  The covariance matrix (\ref{covmat}) is computed for a fixed number $N_{4}$ of 4-simplices constituting the triangulations; thus in line with standard finite-size scaling techniques, we should compute (\ref{covmat}) for a sequence of increasing $N_4$, at each stage finding a best-fit $(i,k)_{N_{4}}$ until the fit stabilizes at some $(i,k)=\lim_{N_4 \to \infty}(i,k)_{N_4}$.  The path integral computed with measure $\left[ \mathcal{D}a \right]_{(i,k)}$ should then represent the continuum limit of CDT, with all modes integrated out except for the scale factor.

An extension of the work presented here is to derive the path integral propagators for all Steigl-Hinterleitner orderings directly rather than in relation to the Laplace-Beltrami case.  While the latter method (that followed in the current paper) may be more efficient for computing a comparison of correlators across operator orderings, a direct derivation of path integral expressions for all Steigl-Hinterleitner orderings is likely to afford conceptual insight into the varying path integral measures.  The aim is to construct such a derivation along the lines of that for the Laplace-Beltrami case using a quantum correction from the Weyl ordering prescription and short-time propagator \cite{DOlivoTorres}.

Further avenues for exploration are comparisons of the CDT data with computations in more sophisticated quantum cosmological models such as Bianchi IX \cite{RyanShepley}.  Allowing the factor ordering of Bianchi IX to vary analogously to the Steigl-Hinterleitner orderings for FRW, it would be interesting to ask whether the same ordering for the kinetic term of the scale factor yields a best fit in both cases.  More generally, consider a sequence of cosmological models such that the degrees of freedom included in a given model are a subset of those in the next model (e.g., FRW and Bianchi IX).  By comparing results from such a sequence of models with computations from a CDT simulation, one could ask whether the best fit factor ordering for a given mode (e.g. the scale factor) remains constant across cosmological models of increasing complexity.  In this way, it would be possible to ask what is the role of factor ordering in quantum cosmological models:  does the best-fit ordering change at each stage, simply compensating for other modes which have been neglected/integrated out?  Alternatively, does the ordering remain stable with the inclusion of new modes, suggesting that it may genuinely reflect qualities of the factor ordering for full general relativity?

To form an estimate of the insights likely to be gained from such a sequence of cosmological models, one could study an analogous sequence of cosmologies in two spacetime dimensions, where CDT's continuum measure is analytically known (see e.g. \cite{AmbjornBook} for coverage of 2d CDT).  This line of inquiry may lead to hypotheses in an important open area:  the nature of four-dimensional CDT's continuum measure.
\\

\textbf{Acknowledgments.}  It is a pleasure to thank Prof. dr. Renate Loll for many helpful discussions, and for insightful comments on the CDT measure.  I would also like to thank Dr. Bianca Dittrich for useful conversations about quantization on a curved background.

\end{document}